\documentclass[12pt]{article}
 \def\cen{\centerline}

\begin{document}

\setlength{\unitlength}{1mm}
%\baselineskip 0.1cm
%\large

%\setcounter{page}{1}
%\pagenumbering{arabic}
 \title{Multidimensional global monopole in presence of Electromagnetic Field}
 \author{\Large $F.Rahaman^*$,P.Ghosh,M.Kalam and S.Mandal}
\date{}
 \maketitle
 \begin{abstract}
                  We study the gravitational properties of a global monopole in
                  $( D = d + 2 )$ dimensional space-time in presence of electromagnetic field.
  \end{abstract}

  %\bigskip
 %\medskip

 \cen{ \bf 1. INTRODUCTION }

 \bigskip
 \medskip
  \footnotetext{ Pacs Nos : 11.27. + d, ; 04.04. -b , 04.50 + h  \\
     \mbox{} \hspace{.2in} Key words and phrases  : Multidimensional global monopole, Electromagnetic field, Test particles.\\
                              $*$Dept.of Mathematics,Jadavpur University,Kolkata-700 032,India\\
                              E-Mail:farook\_rahaman@yahoo.com
                              }

    \mbox{} \hspace{.2in}  There are so many particle physics models solutions that correspond
    to a class of topological defects, which are either stable or long-lived. It is generally
    believed that such topological defects could have been formed naturally during phase
    transitions followed by spontaneously broken symmetries in the early stages of
    the evolution of the Universe. They can be of various types such as monopoles, domain walls,
    strings and their hybrids [1]. Their nature depends on the topology of the vacuum manifold
    of the field theory under consideration. These topological defects have an important role
    in the formation of our universe. Global monopole, which has divergent mass in flat
    space-time, is one of the most important above mentioned defects. At first, Barriola and
    Vilenkin (BV) have studied the properties of the global monopole in curved space-time [2].
    They found a peculiar result: The space-time produced by a global monopole has no Newtonian
    gravitational potential in spite of the geometry produced by this heavy object has a
    non-vanishing curvature. Finding a theory that unifies gravity with other forces in nature
    remains an elusive goal in modern physics. Most efforts in this search have been directed
    at studying theories in which the number of dimensions of space-time is greater than usual
    4-dimensional space-time. For these reasons higher dimensional theory is receiving great
    attention in both cosmological and non-cosmological phenomena
    [3].\\
    It has also been suggested that the experimental detection of the time variation of
    fundamental constants could provide strong evidence of the extra dimension [4]. There has
    been a fairly large amount of discussions on the gravitational field of global monopole
    beginning with the work of BV. Harrari and Luosto[5] , and Shi and Li [5] have shown the
    gravitational potential of global monopole is actually repulsive. Dando and Gregory[5]
    studied global monopole in dilaton gravity. Barros and Romero [5], Banerji et al [5] ,
    E.R.B. de Mello [5] studied global monopole in Brans-Dicke theory. Hao and Li [5] studied
    global monopole in Asymptotically dS / AdS space-time.\\
    In recent past, Banerji et al [6] have obtained a monopole solution in Kaluza-Klein
    space-time, which extends the earlier work of BV to its five dimensional analogue. In a
    recent work, Bronnikov and Meierovich (BM) [7] studied gravitational properties of global
    monopole in $( D = d + 2 )$ dimensional space-time. It is generally believed that during the
    early stage, the Universe could be without geometrical regularity and thermal equilibrium.
    It is well known that the magnetic field has a significant role at the cosmological scale
    and is present in galactic and inter galactic spaces. Melvin [8] has pointed out that during
    a large part of the history of evolution matter has highly ionized, smoothly coupled with
    the field and subsequently formation neutral matter during expansion of the Universe. In
    the present work, we study gravitational properties of global monopole in $( D = d + 2
    )$
    dimensional space-time in presence of electromagnetic field.

    \bigskip
   \medskip
    \cen{ \bf 2. Field Equations and their integrals:  }
    \bigskip
    \medskip

       In this section we closely follow the formalism of BM and take a D dimensional
       space-time with the structure $R^1 X S^1 X S^d ( d = D - 2 ),$ where $S^1$ is the range of
       the radial coordinate $r$ and $R^1$ is the time axis.We take a static, spherically symmetric
       metric in $D = d + 2$ dimension as

\begin{equation}
                ds^2 = e^{2\gamma}dt^2 - e^{2\mu}dr^2-r^2
                d\Omega_d ^2
         \label{Eq1}
          \end{equation}

Here $ d\Omega_d ^2$ is a linear element on a d-dimensional unit
sphere, parametrized by the angles $\phi_1,
\phi_2,......,\phi_d:$ \\

$d\Omega_d ^2= d\phi_d^2 + \sin_2 \phi_d [d\phi_{d-1}^2 + \sin_2
\phi_{d-1}[d\phi_{d-2}^2 + .........+ \sin_2 \phi_3(d\phi_2^2 +
\sin_2 \phi_2  d\phi_1^2).......]] $ \\

  A global monopole with a non-zero topological charge can be constructed with a multiplet
  of real scalar fields $\Phi^a ( a = 1,2,....., d +1) $comprising a hedgehog configuration in
  $( d + 1)$ spatial dimension [7]\\
\begin{equation}
        \Phi^a = \eta f(r) n^a (\phi_1, \phi_2,...... , \phi_d )
         \label{Eq2}
          \end{equation}
where $n^a(\phi_1, \phi_2,.......... , \phi_d )$is a unit vector
$( n^a n^a = 1 ) $ in the $d-dimensional$ Euclidean target space,
with the components \\

$ n^{d+1} = \cos \phi_d ,$ \\
$ n^{d} = \sin \phi_d \cos \phi_{d-1} ,$\\
$ n^{d-1} = \sin \phi_d \sin \phi_{d-1}\cos \phi_{d-2} ,$\\
....................................................... \\
....................................................... \\
$ n^{d-k} = \sin \phi_d \sin \phi_{d-1}.........\sin \phi_{d-k}\cos \phi_{d-k-1} ,$\\
.........................................................\\
.........................................................\\
$ n^{2} = \sin \phi_d \sin \phi_{d-1}.........\sin \phi_{2}\cos \phi_{1} ,$\\
$ n^{1} = \sin \phi_d \sin \phi_{d-1}.........\sin \phi_{2}\sin \phi_{1} .$\\
The Lagrangian of a multi-dimensional global monopole in general
relativity has the form\\
\begin{equation}
         L=\frac{1}{2}\partial_{\mu}\Phi^{a}\partial^{\mu}\Phi^{a}-
              V(\phi)+ \frac{1}{16\pi G} R
         \label{Eq3}
          \end{equation}
where $R$ is the scalar curvature in the $D -dimensional$
gravitational constant and $V (\phi)$ is a symmetry breaking
potential depends on $\phi = \sqrt{\Phi^a \Phi^a}$ and it is
natural to choose $V$ as the Mexican-hat potential
\begin{equation}
              V(\phi)= \frac{1}{4}\lambda (\Phi^a \Phi^a -
              \eta^2)^2 =\frac{1}{4}\lambda \eta^4 ( f^2 - 1)^2
         \label{Eq4}
          \end{equation}
The model has a global $S0( d + 1)$ symmetry, which can be
spontaneously broken to $S0(d);\eta^{\frac{2}{d}}$ is the energy
of symmetry breaking.\\
One can find easily the components of energy momentum tensor
$(T_b^a )$ by using the Lagrangian (3) and metric (1).\\
The non-zero components of $^*T_b^a ( = T_b^a - \frac{1}{d}T
\delta_b^a )$ are \\

 \cen{$        ^*T_t^t = -\frac{2}{d} V $}

\medskip
 \cen{$        ^*T_r^r = -e^{-2\mu}(f^\prime)^2 - \frac{2}{d} V $ }

\begin{equation}
         ^*T_2^2 = ............ = ^*T_{d+1}^{d+1} = - \frac{\eta^2
         f^2}{r^2} - \frac{2}{d} V
         \label{Eq5}
          \end{equation}

The scalar field equation $\nabla^\alpha \nabla_\alpha \Phi^a  +
\frac{\partial V}{\partial \Phi^a } = 0 $ ( $ \nabla^\alpha
\nabla_\alpha $ is the D'Alembert operator) in flat space-time is
\begin{equation}
         (r^d \phi ^ \prime)^\prime - dr^{d-2} \phi =r^d (\frac{\partial V}{\partial
         \phi})
         \label{Eq6}
          \end{equation}
   where $ \phi =\sqrt{\Phi^a \Phi^a}$

A special solution of this equation is [7]\\
\begin{equation}
         f= 1- \frac{d}{2\lambda \eta^2 r^2}
         \label{Eq7}
          \end{equation}
[ $^{'\prime '}$ refers to differentiation  with respect to radial
coordinate ]\\

   It can be shown that in flat space the mass of the monopole core,
   $M_{core}\sim \lambda^{-\frac{1}{d}}\eta ^ {\frac{2}{d}}$ . Thus if $ \eta \ll m_p $
   where $m_p$ is the plank mass, it is evident that we can still apply the flat space
   approximation of $M_{core}$ . This follows from the fact that in this case the gravity
   would not much influence on monopole structure.\\
   Outside the core, however, one can assume  $ f= 1- \frac{d}{2\lambda \eta^2
   r^2} $ .\\
   The stress energy tensor of the electromagnetic field has the expression
\begin{equation}
        E_{ab}=[ F_a^c F_{bc} - \frac{1}{4}g_{ab}F_{cd}F^{cd} ]
         \label{Eq8}
          \end{equation}
  where $F_{ab}$ is the electromagnetic field tensor. \\

Following Xu Dianyan [9], one can write the Maxwell equations as
follows:
\begin{equation}
        {F_b^a}_{;b} = 0
         \label{Eq9}
          \end{equation}
\begin{equation}
        F_{ab;c} +  F_{bc;a} +  F_{ca;b} = 0
         \label{Eq10}
          \end{equation}
\begin{equation}
        F_{ab} = A_{a,b} -  A_{b,a}
         \label{Eq11}
          \end{equation}
where $A_a$ is the electromagnetic vector potential.\\

The components of the electromagnetic field tensor, not equal to
zero, are [9]
\begin{equation}
        F_{12} = -F_{21} =\frac{q}{r^d}
         \label{Eq12}
          \end{equation}
and the corresponding vector potential is
\begin{equation}
        A_1 = \frac{q}{(d-1)r^{d-1}}
         \label{Eq13}
          \end{equation}
where $q$ is the total charge within the monopole core.\\

[ here the suffix 1 refers to time and 2 to radial coordinate ]\\

The Einstein field equation \\

 \cen{$  R_b^a = - 8\pi G [^*T_b^a + ^*E_b^a  ]$}

 where \\
\cen{$^*T_b^a  = T_b^a - \frac{1}{d}T \delta_b^a $ }
and \\
\cen{$^*E_b^a  = E_b^a - \frac{1}{d}E \delta_b^a $}
are \\
\begin{equation}
  e^{-2\mu}[\gamma^{\prime\prime} +
  \frac{1}{2}(\gamma^\prime)^2-\gamma^\prime \mu^\prime + \frac{d
  \gamma^\prime}{r}] = - 8 \pi G [- \frac{d}{2\lambda r^4}+ (1 -
  \frac{1}{d})q^2 r^{-2d}]
         \label{Eq14}
          \end{equation}
\begin{equation}
  e^{-2\mu}[\gamma^{\prime\prime} +
  \frac{1}{2}(\gamma^\prime)^2-\gamma^\prime \mu^\prime - \frac{d
  \mu^\prime}{r}] = - 8 \pi G [- \frac{d}{2\lambda r^4}+ (1 -
  \frac{1}{d})q^2 r^{-2d}]
         \label{Eq15}
          \end{equation}
\begin{equation}
  e^{-2\mu}[\frac{(d-1)}{r^2}+\frac{\gamma^\prime -
  \mu^\prime}{r}] - \frac{(d-1)}{r^2}
   = - 8 \pi G [\frac{d}{2\lambda r^4}- \frac{\eta^2}{r^2} -
  \frac{1}{d}q^2 r^{-2d}]
         \label{Eq16}
          \end{equation}
For the reason of economy of space we will skip all mathematical
details and give the final result as \\
for $ d \neq 3 $ we have
\begin{equation}
  e^{2\gamma}=e^{-2\mu}=1 + \frac{1}{(d-1)}8 \pi G \eta^2- M
  r^{1-d}- \frac{d}{\lambda(d-3)} \frac{4\pi G }{r^2} -
  \frac{d}{(d-1)} 8 \pi G q^2 r^{2-2d}
         \label{Eq17}
          \end{equation}
for $ d = 3 $ we have
\begin{equation}
  e^{2\gamma}=e^{-2\mu}=1 + 4 \pi G \eta^2- M r^{-2}- \frac{12 \pi G}{\lambda r^2} \ln r
  + \frac{4}{3} \frac{\pi G q^2}{r^4}
         \label{Eq18}
          \end{equation}

 \bigskip
 \medskip

 \cen{ \bf 3.  Motion of test particles:   }

 \bigskip
 \medskip
          Let us consider a test particle having mass 'm' moving in the gravitational field
          of a D-dimensional monopole described by the metric ansatz
          (1).\\
          So the H-J equation for the test particle is [10]

\begin{equation}
   g^{ik}\frac{\partial S}{\partial x^i} \frac{\partial S}{\partial
   x^k}+ m^2 = 0
         \label{Eq19}
          \end{equation}
   where $ g_{ik}$ are the classical background  field (1) and S is the standard Hamilton's
   characteristic function . For the metric (1) the explicit form of H-J equation (19) is  [10]
\begin{equation}
  - \frac{1}{g}(\frac{\partial S}{\partial t})^2 + g(\frac{\partial S}{\partial
  r})^2+ \frac{1}{r^2}[ (\frac{\partial S}{\partial x_1})^2+(\frac{\partial S}{\partial x_2})^2
  + ............... + (\frac{\partial S}{\partial x_{D-2}})^2] +  m^2 = 0
         \label{Eq20}
          \end{equation}
 with $x_1, x_2 ,.......,x_{D - 2}$ are the independent coordinates on the surface of the unit
$( D - 2 )$ sphere such that \\
\begin{eqnarray*} \\&&d \Omega_{D-2}^2 = dx_1^2 + dx_2^2 +
.............+dx_{D-2}^2\\&& \equiv d\theta_1^2 +\sin ^2 \theta_1
d\theta_2^2+ ............... + \\&&\sin ^2 \theta_1\sin ^2
\theta_2.............\sin ^2 \theta_{D-3}d\theta_{D-2}^2
\end{eqnarray*}

and $ g $ is given in (17 ) or (18).\\

  In order to solve this partial differential equation, let us choose the $H-J$ function $ S $ as
   \begin{equation}
       S = - E.t + S_1(r) + p_1.x_1 + p_2.x_2 +........... + p_{D - 2}.x_{D -
       2}
         \label{Eq21}
          \end{equation}
 where $E$ is identified as the energy of the particle and $p_1, p_2 ,......, p_{D - 2} $
 are the momentum of the particle along different axes on the $( D - 2 )$ sphere with
 $  p = \sqrt{ p_1^2 + p_2^2 + ...... + p_{D - 2}^2} $, as the resulting momentum of the particle
 .\\

  If we substitute the ansatz (21) for $S$ in the $H-J$ equation (20) then we get the
  following expression for the unknown function  $S_1$:

\begin{equation}
         S_1(r)  =  \epsilon \int \sqrt{\frac{E^2}{g^2} - \frac{m^2}{g} - \frac{p^2}{r^2 g}} dr
         \label{Eq22}
          \end{equation}
where $\epsilon = \pm  1 $, the sign changing whenever  $r$ passes
through a zero of the integral (22). \\

To determine the trajectory of the particle following $H-J$
method, we consider [10] \\

  $ \frac{\partial S}{\partial E} $ = constant;  $\frac{\partial S}{\partial p_i}$ = constant ;
   [ $i = 1,2,........,(D - 2 ) $] .\\
    ( we have chosen the constants to be zero without any loss of generality
    .)\\

  Hence we get
\begin{equation}
         t = \epsilon \int \frac{E/g^2}{\sqrt{\frac{E^2}{g^2} - \frac{m^2}{g} - \frac{p^2}{r^2 g}}} dr
         \label{Eq23}
          \end{equation}
\begin{equation}
         x_i = \epsilon \int \frac{p_i/r^2g}{\sqrt{\frac{E^2}{g^2} - \frac{m^2}{g} - \frac{p^2}{r^2 g}}} dr
         \label{Eq24}
          \end{equation}
  From (23), the radial velocity of the particle  is
\begin{equation}
         \frac{dr}{dt} =  \frac{\sqrt{\frac{E^2}{g^2} - \frac{m^2}{g} - \frac{p^2}{r^2 g}}}{E/g^2}
         \label{Eq25}
          \end{equation}
 The turning points of the trajectory are given by $\frac{dr}{dt} = 0 $ and as a consequence
 the potential curve are
\begin{equation}
         \frac{E}{m} = \sqrt{g (\frac{p^2}{m^2r^2} + 1)} \equiv V
         (r)
         \label{Eq26}
          \end{equation}
In a stationary system $ E $ i.e. $ V(r)$ must have an extremal
value. Hence the value of $r$ for which energy attains it
extremal value is given by the equation
\begin{equation}
         \frac{dV}{dr} =   0
         \label{Eq27}
          \end{equation}
Hence we get the following equation as \\
for $ d \neq 3 $ we have
\begin{eqnarray*}
  \\
  &&r^{2d-2}[2p^2(1 + \frac{8\pi G\eta^2}{d-1})- \frac{8\pi Gm^2d}{\lambda(d-1)}]\\
  &&-M
  m^2(d-1)r^{d+1}-\frac{16\pi Gp^2d}{\lambda (d-3)}
  r^{2d-4}\\
  &&-M p^2(d+1)r^{d-1}-\frac{16\pi Gq^2m^2r^2}{\lambda}-\frac{16\pi
   Gq^2p^2}{d-1}=0
      \label{Eq28}
          \end{eqnarray*}
This is an algebraic equation with negative last term provided \\

\cen{$2p^2 ( 1 + \frac{8 \pi G \eta^2}{d-1}) > \frac{8 \pi G
m^2d}{\lambda(d-1)}$ .}

This equation must have at least one real positive root. So the
bound orbits are possible for the test particle i.e. particle can
be trapped by global monopole.\\

for $ d = 3 $ we have
\begin{eqnarray*}
   \\
  &&\frac{24 \pi G}{\lambda} \ln r ( p^2 m^2 + m^2r^4 + p^2r^2 )
  +
  4 \pi G p^2q^2 \\
  &&= [2p^2(1+4 \pi G \eta^2)+ \frac{12 \pi G m^2}{\lambda}] r^4 +\\
  && [\frac{12 \pi G p^2}{\lambda}- \frac{4}{3}\pi G m^2q^2]r^2
         \label{Eq29}
          \end{eqnarray*}
This equation has real solutions provided $\frac{9p^2}{\lambda}
> m^2q^2 $ , so orbit of a massive test particle must be trapped by
monopole.\\

 \bigskip
 \medskip
 \cen{ \bf 4. Concluding Remarks:  }

       In this work, we have studied the gravitational field of global monopole in higher
 dimensional space-time with topology is $ R^1 X S^1 X S^d $ in presence of electromagnetic
 field. We have found two sets of solutions: one for $ d \neq 3 $ and other for $ d = 3 $. It is shown
 that higher dimensional global monopole exerts attractive gravitational force on test particles
 provided some restrictions ( including the dimension of the space-time ) to be imposed.
 Thus one can conclude that dimensional factor ( to some extend ) effects the motion of the
 test particle.\\

        { \bf Acknowledgement }

        One of the authors, F.R is thankful to IUCAA for providing the research facility. \\

%\begin{figure}[p]
%\includegraphics*[450,350]{fig1.bmp}
%\caption{Variation of deflection of the circular plate}
%\end{figure}

\end{document}